\documentclass[letterpaper,12pt]{article}   
\usepackage{osajnl} 
\usepackage[draft]{hyperref} 
\usepackage[]{aas_macros}

\begin{document}

\title{Robust determination of optical path difference: fringe tracking at the IOTA  interferometer}

\author{Ettore Pedretti, Wesley A. Traub, John D. Monnier, Rafael Millan--Gabet,  Nathaniel P. Carleton, F. Peter Schloerb, Michael K. Brewer, Jean--Philippe Berger, Marc G. Lacasse and Sam Ragland.}
\email{epedretti@umich.edu}
\affil{Ettore Pedretti, Wesley A. Traub, Sam Ragland, Nathaniel P. Carleton,  Marc G. Lacasse are with the Smithsonian Astrophysical Observatory, 60 Garden Street,Cambridge, MA, 02138, USA. 
John Monnier and Ettore Pedretti are at the University of Michigan, Ann Arbor, MI 48109, USA. 
Rafael Millan--Gabet is at the Michelson Science Center, California Institute of Technology, Pasadena, CA 91125, USA.
F. Peter Schloerb and Michael. K. Brewer are with the University of Massachusetts, Amherst, MA 01003, USA. 

Jean--Philippe Berger is with the Laboratoire d' Astrophysique de Grenoble, BP 53 X, Grenoble Cedex 9, France.}

\begin{abstract}
We describe  the fringe  packet tracking system  used to  equalise the
optical path  lengths at the  Infrared Optical Telescope  Array (IOTA)
interferometer. The  measurement of closure  phases requires obtaining
fringes on three baselines  simultaneously. This is accomplished using
an algorithm based on  double Fourier interferometry for obtaining the
wavelength--dependent phase of the  fringes and a group delay tracking
algorithm  for determining  the  position of  the  fringe packet.  The
comparison between  data acquired with  and without the  fringe packet
tracker  shows  about  a  factor  3  reduction of  the  error  on  the
closure--phase measurement. The fringe packet tracker has been able so
far to track  fringes of signal--to--noise as low as  1.8 for stars as
faint as $m_H=7.0$.
\end{abstract}


\maketitle 

\section{Introduction}

Observations     performed    with     long--baseline    ground--based
optical/infrared   interferometers  are   strongly  affected   by  the
turbulent atmosphere. Turbulence can  reduce the visibility of fringes
in many ways as described by Porro~et~al\cite{1999ApOpt..38.6055P} for
pupil--plane      (or      coaxial)      beam     combination      and
Thureau\cite{2001PhDT.......161B}  for image--plane  beam combination.
Turbulence randomly modulates the phases of the fringes which can then
become  unusable  for  image   reconstruction.  Using  three  or  more
telescopes removes the atmospheric  phase contamination.  This is done
through    the   closure--phase    technique   pioneered    in   radio
astronomy\cite{1958MNRAS.118..276J}  and   only  recently  applied  to
long--baseline  optical  interferometry\cite{1996A&A...306L..13B}. The
necessary condition  for obtaining meaningful  closure--phases is that
the  three  fringe  packets  must  be detected  within  the  coherence
time. This is achieved by keeping the optical path difference (OPD) to
a minimum.

Prior to 2002, IOTA had only two telescopes and a single baseline, and
the fringes were usually kept inside the scan interval manually by the
observers. The installation of the third telescope at IOTA required an
increase in the level of  automation in the instrument, because manual
tracking  is  not  practical   with  three  baselines  to  adjust.  In
particular the  requirement to measure  closure--phases necessitated a
system capable of keeping the fringe packets in the centre of the scan
using the existing hardware  dedicated to acquiring data. Fringes must
be  acquired  in  the  same  coherence  time in  order  to  measure  a
closure--phase.  This  is  especially  important  at  IOTA  where  the
bandwidths are  frequently relatively large (15--25\%)  and the fringe
packets quite narrow (3--6 fringes).

We present here a simple and fast algorithm, which uses double Fourier
interferometry (DFI)  \cite{1988A&A...195..350M} to extract wavelength
dependent information  from the fringe packet and  calculate its group
delay.   The   same   algorithm   was  independently   discovered   by
Tubbs\cite{tubbs98} and written  as a final-year undergraduate project
but never published. In this report Tubbs tested the algorithm through
simulated data  and data obtained  from the COAST  interferometer, but
did  not implement a  working fringe--packet  tracker. We  only became
aware of this work after our algorithm was routinely used at the IOTA.

The time--domain interferograms, recorded  by an infrared detector for
the purpose  of measuring the physical parameters  of stellar objects,
are used in order to track the position of fringes. A fringe packet is
generated   by  pupil-plane   interference  of   starlight   from  two
telescopes,   where   the   optical   path   difference   around   the
zero-path-difference point  is changed linearly with  time.  Each pair
of  the  three  telescopes  generates two  such  interferograms,  with
complementary intensities.   The goal is to control  the optical paths
such that  these interferograms  occur nearly simultaneously,  so that
the  closure  phase  can be  measured  from  them.   To do  this,  the
algorithm in the  present paper is applied to  each fringe packet, and
the  output of  the algorithm  is used  to control  the  optical paths
before  the   next  interferogram  is   produced.   Up  to   about  10
interferograms are  generated per  second, using triangular  sweeps of
the optical paths.  No additional hardware is required.

The  application  of  the  algorithm  is  not  restricted  to  stellar
interferometry  but can  be  applied to  all  cases where  broad--band
interference  fringes must  be tracked.   Other algorithms  for fringe
packet tracking  which have  been tested at  IOTA are the  subjects of
separate                                                   publications
\cite{1999SPIE.3749..691W,2000SPIE.4006..506M,2003SPIE.4838..956T,2005EURASIP..Q2W,2005SPIE..5491..323W}.

\section{The Instrument}

IOTA  is  a  long--baseline  optical  interferometer  located  at  the
Smithsonian  Institution's Whipple Observatory  on Mount  Hopkins, AZ,
comprising three 45--cm diameter telescopes which can be positioned at
17 stations on an L--shaped track,  where the arms are 15~m toward the
south--east  and 35~m toward  the north--east.   IOTA operated  with 2
telescopes from 1995--2003, and 3 telescopes since February 2002.  The
interferometer\cite{2000SPIE.4006..715T,2003SPIE.4838...45T}, has been
used      as      a       testbed      for      new      cutting--edge
technologies\cite{2001A&A...376L..31B,2002SPIE.4838..1127,2004PASP..116..377P},
and   has    produced   astronomy   results    in   the   3--telescope
configuration\cite{2002SPIE.4838..379,2002SPIE.4838..1068,2002SPIE.4838..202,2002SPIE.4838..181,2004ApJ...602L..57M}.

The three  beams arriving from  the vacuum delay--line tank  hit three
dichroic mirrors  which separate the  visible and infrared  light. The
visible beam  continues toward the  star tracker servo system.  In the
implementation discussed  here, the infrared beam  is reflected toward
three flat mirrors and then  three off--axis parabolas which focus the
three  beams  on  three  single--mode  (H--band)  fibers  feeding  the
IONIC--3T integrated--optics beam--combiner\cite{2003SPIE.4838.1099B}.

Interference  is  achieved  inside  the integrated  optics  component,
resulting  in  three output  pairs  $\pi $  radians  out  of phase  in
intensity.  The interference  fringes are  recorded while  two  of the
dichroics are  piezo--driven to scan a  path of about  +50~$\mu $m and
-50~$\mu $m, respectively, in order  to scan through the fringe packet
in the  three beams. The  six combined beams  are then focused  on six
separate  pixels  of  the PICNIC  array\cite{2004PASP..116..377P}  and
recorded as time series for  science measurement. The same time series
is used by the  fringe--packet tracker. The path difference calculated
by the  packet--tracker is fed  back to the  piezo--scanning dichroics
for a fast  tracking response.  The piezo scanners  are off--loaded of
their additional  offsets every second,  when a fraction of  the error
signal  is sent to  the short  delay lines  which are  responsible for
tracking the geometric delay caused by the rotation of the Earth.

\section{Calculating the Fringe Position}
\subsection{Tracking the Fringe Packet Using Double Fourier Interferometry}

Our  method of  fringe packet  tracking at  IOTA calculates  the group
delay  of fringes  dispersed with  DFI, which  is used  to  obtain the
wavelength dependent  phase from  the fringe packet.  This is  done by
scanning the  fringe packet over  an interval greater than  the packet
length,  where the  spectral resolution  is proportional  to  the scan
length.  The group  delay tracking  (GDT) method  has been  applied to
interferometry   since  the   very  beginning   of  the   field,  when
Michelson\cite{1921ApJ....53..249M}  used a  prism for  dispersing and
acquiring       fringes        visually       at       the       20-ft
interferometer.   Labeyrie\cite{1975ApJ...196L..71L}  used   the  same
system  and   demonstrated  fringe  acquisition   on  a  two-telescope
interferometer.

Several  systems have  been proposed  since then,  for  correcting the
optical path\cite{1989SPIE.1130..109V,1994SPIE.2200..204R}.  GDT (also
called  dispersed   fringe  tracking  when  applied   to  image  plane
interferometry)  has been  routinely used  at  several interferometric
facilities\cite{1994SPIE.2200..222R,1995OSAJ...12..366L,1996ApOpt..35.3002K}.
In fact,  at IOTA,  GDT was  selected as the  original method  of path
difference               monitoring               in               the
visible\cite{1987iia..conf..129N,1988hrii.conf..947T,1990OSAJ....7.1779T,1990SPIE.1237..145T},
but the system was set aside in favor of making infrared observations.

We will perform now a derivation of the algorithm for a two--telescope
interferometer which combines  the light in the pupil  plane through a
beam--splitter. With this setup the light intensity $I(\xi )$ from the
two  complementary  outputs  of  the  beam--splitter  is  measured  by
square--law  detectors:  
{\begin{equation}  
I(\xi )=I_{t}\left\{  1\pm V\frac{\sin    \left[\pi    \left(\xi    -\xi    _{0}\right)\Delta f\right]}{\pi   \left(\xi    -\xi   _{0}\right)\Delta   f}\sin \left[2\pi  \left(\xi  -\xi _{0}\right)f_{0}+\phi  \right]\right\},\label{eq:intensityNoBkg}
\end{equation} 
where $I_{t}$  is the mean intensity  which becomes split  by the beam
combiner ($\pm$ is for complementary  outputs) and is modulated by the
interferometric signal, function  of the optical path $\xi  $.  $V$ is
the contrast of  the fringes, $\Delta f$ (cm$^{-1}$)  the bandwidth of
the  ideal  (rectangular)  spectral  filter, $f_{0}$  (cm$^{-1}$)  the
frequency of the fringe, $\phi$  (rad) a generic phase, and $\xi _{0}$
(cm)  is the  fringe--packet centre,  which varies  from scan  to scan
owing to  atmospheric path  fluctuation. This is  the actual  value we
want to calculate in order to correct the optical path.

In practice we  are dealing with discrete intensities,  expressed as a
finite series of data,  so we can rewrite (\ref{eq:intensityNoBkg}) as
the discrete function $n(j)$, which is the detector count.
\begin{equation}
n(j)=n_0\left\{ 1\pm V \frac{\sin \left[\pi \left(j-J_{0}\right)\Delta m\right]}{\pi \left(j-J_{0}\right)\Delta m}\sin \left[2\pi \left(j-J_{0}\right)m_{0}+\phi \right]  \right\} +\epsilon_j \label{discrInt}
\end{equation}
Here $n_0$ is the mean number of electrons per channel, and $V$ is the
fringe visibility.   The sample number j  ranges from 1  to $N$, where
$N$ is the  total number of samples; typically $N  = 256$.  The centre
of the  fringe packet  is at  sample number $J_0$.   If the  high- and
low-wavenumber limits of the spectral  filter are written as $m_h$ and
$m_l$ (waves  per sample), then the filter  full-width at half-maximum
is  $\Delta m  = m_h  - m_l$,  and the  filter centre  is $m_0  = (m_h
+m_l)/2$, both in  units of waves (or fringes)  per sample.  The phase
$\phi$ (radians)  is an offset  between the sinusoid carrier  wave and
the  centre of the  sinc envelope.   The $\epsilon_j$  term represents
additive noise, the sum of electron counting statistics, detector read
noise, and random atmospheric scintillation. The goal is to extract an
estimate of $J_0$ from the ensemble of $N$ data points.

The first step is to calculate  the fast Fourier transform of the data
string $n(j)$, and scaling the result so as to eliminate uninteresting
factors.  The result is a  sequence of $N/2$ complex numbers which can
be written as
\begin{equation}
    \tilde{n}\left(m_0 \right) = \exp \left(-i 2 \pi m_0 J_0 \right) + \eta_{m_0}  \label{discrFFT}       
\end{equation}
where  $\eta_{m_0}$  is  the   scaled  transform  of  the  noise  term
$\epsilon_j$.

The  second step  is  to select  two  of the  $\tilde{n}$ values,  say
$\tilde{n}(m_0)$ and  $\tilde{n}(m_0 +  \Delta m_{12})$, and  form the
cross-spectrum product of the first  with the complex conjugate of the
second, i.e.,
\begin{equation}
X\left(m_0, m_0+\Delta m_{12}\right) = \tilde{n}(m_0) \cdot \tilde{n}\left(m_0 + \Delta m_{12}\right)^{*}. \label{xspec}
\end{equation}

As  can be  seen from  (\ref{discrFFT}), the  cross-spectrum  $X$ will
generate  a  complex number  (plus  noise)  whose  phase contains  the
unknown quantity $J_0$, multiplied by known terms.  The additive noise
would  degrade our  estimate of  $J_0$ from  the cross  spectrum term,
however  adding several  such terms  will improve  the signal-to-noise
ratio of the resulting complex term\cite{1988PhDT.......202B}, and likewise improve our estimate
of $J_0$.

The   third  step   is   to  calculate   the  average   cross-spectrum
$\overline{X}$, as for example in
\begin{equation}
\overline{X}=\sum\limits _{m_0=m_{l}}^{m_h} \frac{X(m_0,m_0+\Delta m_{12})}{(m_h -m_l +1)}= \exp \left(-i2\pi \Delta_{m12}J_{0}\right)+\overline{\eta} , \label{eq:discXspec}
\end{equation}
where $\overline{\eta}$ is the new noise term, presumably smaller than
the  root-mean-square of  $\eta_{m_0}$ by  a  factor on  the order  of
$\sqrt{(m_h-m_{l+1})}$.   The  limits  on  the sum  are  only  suggested
values,  and could  be changed  so as  to improve  the signal-to-noise
ratio if applicable.

(We note  that the  cross-spectrum is also  used in  the Knox-Thompson
algorithm\cite{1974ApJ...193L..45K}      for      recovering      near
diffraction-limited images of  stellar objects from snapshots degraded
by atmospheric seeing.)

The estimated fringe packet position $<J_0>$ can now be recovered from the average cross-spectrum using
\begin{equation}
<J_{0}>=\frac{\arg \left(\overline{X}\right)}{2\pi \Delta m_{12}}.\label{opd3}
\end{equation}
Here the estimated packet center $<J_0>$ is measured in units of
sample numbers from the first point in the interferogram.  In
practice we subtract $N/2$ from this value, and send the resulting
value (suitably scaled) as an error signal to the servo system,
such that the starting point of the next scan is adjusted
accordingly.

We note that any value of $\Delta m_{12}$ could be used, but we
suggest that $\Delta m_{12} = 1$ is optimum in the sense that it is least
likely to produce a complex $S$ or $\overline{S}$ that is biased by
wrap-around effects, which will occur if the phase shift between
adjacent values of $\tilde{n}$ are separated by more than $\pi$.
Hereafter we drop the expectation value brackets, and write $J_0$ for
$<J_0>$ for notational simplicity.

\subsection{Algorithm summary}
\begin{enumerate}
\item  The interference fringes are recorded while two of the optical paths are piezo--driven
to scan a path of about +50~$\mu $m and -50~$\mu $m, respectively. The time series $n_{j}$ is recorded by the infrared camera for the three pairs of beams. 
\item The fast Fourier transform  $\tilde{n}(m_0)$ of the time series $n_{j}$ is computed separately for the three pairs. 
\item The cross--spectrum $X$ is calculated  for the 3 beams by multiplying $\tilde{n}(m_0)$ by its complex conjugated $\tilde{n}^*(m_0+1)$ shifted by one sample. 
\item The average cross--spectrum $\overline{X}$   is computed in correspondence of the fringe peak in the cross--power--spectrum, in order to average complex vector with higher signal--to--noise.
\item{ The position of the fringe packets is obtained from the phase of the average vector $\overline{X}$. using Eqn.~\ref{opd3}}
\end{enumerate}

\subsection{\label{subsec:Baseline-bootstrapping}Baseline Bootstrapping}

With baseline bootstrapping\cite{1988eso.conf..565N,1998ApJ...496..550A} we are capable of blind-tracking fringes
on a baseline when the signal-to-noise of the fringes is too low,
provided that good signal-to-noise is available on the other two baselines.
For this reason, we calculate the optical path $J_{0}$ for three
baselines even if we correct the path for two baselines. We can express
one optical path as the weighted average of the other two optical
paths, the weights being equal to the SNR for the fringes obtained
on those baselines. The signal-to-noise is calculated from the cross
spectrum $\left|X\right|^{2}$, dividing the
averaged power inside the fringe peak by the averaged power outside
the fringe peak. We then observe that the optical path, in a closed
loop must be equal to zero:

\begin{equation}
J_{0}1+J_{0}2+J_{0}3=0\label{eqnBoost1}\end{equation}
where $J_{0}$1 and $J_{0}2$ are the optical path where the servo
loop is acting, while $J_{0}3$ is the reference optical path. To
the path $J_{0}1$, $J_{0}2$ and $J_{0}3$ are associated the weights
$w1$, $w2$ and $w3$ respectively. We have two values for each optical
path. One is the value obtained directly on that baseline (for example
$J_{0}$1 with weight $w1$), the other is the value calculated from
the linear combination of the other two baselines (for example $J_{0}1'=-J_{0}2-J_{0}3$
with weight $w1'=(w2w3)/(w2+w3)$). The weighted average of $J_{0}$
and $J_{0}'$ can then be written as\cite{1992drea.book.....B}:\begin{equation}
\overline{J_{0}1}=\frac{w1J_{0}1+w1'J_{0}1'}{w1+w1'}\label{eqnBoost2}\end{equation}
where $\overline{J_{0}1}$ is the weighted-averaged path difference.
Similarly for $\overline{J_{0}2}$:\begin{equation}
\begin{array}{lcl}
 \overline{J_{0}2} & = & \frac{w2J_{0}2+w2'J_{0}2'}{w2+w2'}\\
 J_{0}2' & = & -J_{0}1-J_{0}3\\
 w2' & = & \frac{w1w3}{w1+w3}\end{array}
\label{eqnBoost3}\end{equation}

The advantage of using a weighting system is that we do not have to select the best baseline values a priori, but rather the weighting allows them to be selected automatically.

\section{\label{subsec:Simulations}Simulations}
The linearity of the fringe packet tracker (FPT) algorithm was tested 
through simulations.  This simulation
was performed in presence of photon and detector noise, with an average
number of 300 photoelectron/sample for the fringe intensity (Poisson distributed), a fringe
visibility of one and an additional 12-electron mean of Gaussian noise for the
detector. 

It was found that the algorithm is capable of
detecting the sign of the correction, necessary to bring back the
fringes to the centre of the scan even if the main part of the fringe
packet is outside the current scan. The necessary condition is to
have enough signal-to noise so that the side lobes are detectable.
This behavior is shown in the graph of Fig.~\ref{fig:fringe Recover}. 

\begin{figure}[!h]
\includegraphics[  scale=0.6,angle=90]{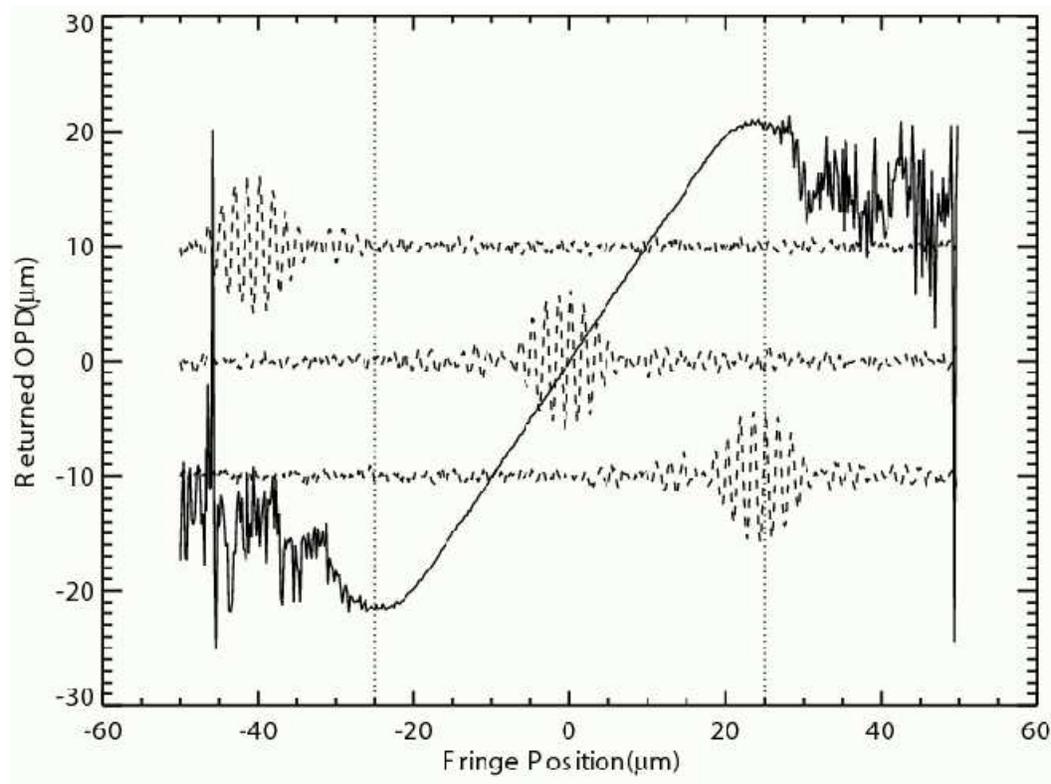}
\caption[The optical path returned by the algorithm.]{\label{fig:fringe Recover}The optical path returned by the algorithm
with respect to the fringe position (continuous-line plot). The two vertical dotted-lines represent the interval inside of which
the fringes are sampled, for a scan length of $50~\mu m$. Three fringe packets (dashed-lines) mark
three representative  positions on the plot: outside the scan range (top), at the
centre of the scan (centre) and half outside the scan range (bottom).
Note that only the side lobes are visible in the scan range of the
plot at the top. }
\end{figure}

The returned position depends linearly on the fringe
positions when the packet is inside the scan but it is 
non-linear when the packet is outside the scan. Nevertheless the information
of the side lobes can still be used to bring the fringes back to the
centre of the scan. The algorithm is sensitive to the slope of the
phase across the bandwidth and the sign of the correction is preserved.
The algorithm starts to fail at the limit of the range where the signal
is too low and the side lobes not detectable for the given integration
time.
The returned position depends linearly on the fringe positions when
the packet is inside the scan and becomes strongly non-linear (but
with the correct sign) when the packet is outside the scan.

\begin{figure}[!h]
\begin{tabular}{cc}
\begin{minipage}[b]{.42\textwidth}
\includegraphics[scale=0.33, angle=90]{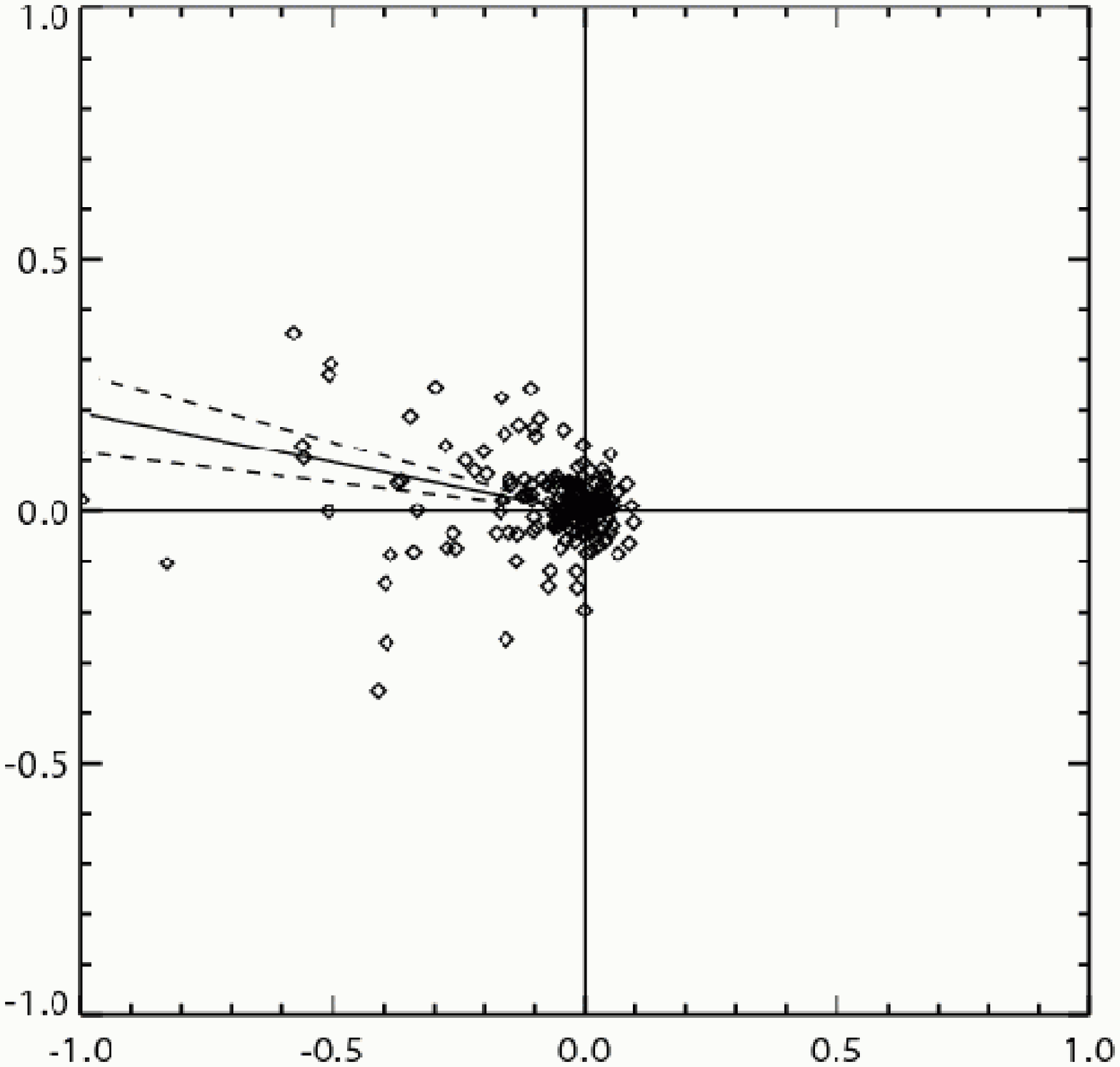} 
\end{minipage} &
\begin{minipage}[b]{.42\textwidth}
\includegraphics[scale=0.33,angle=90]{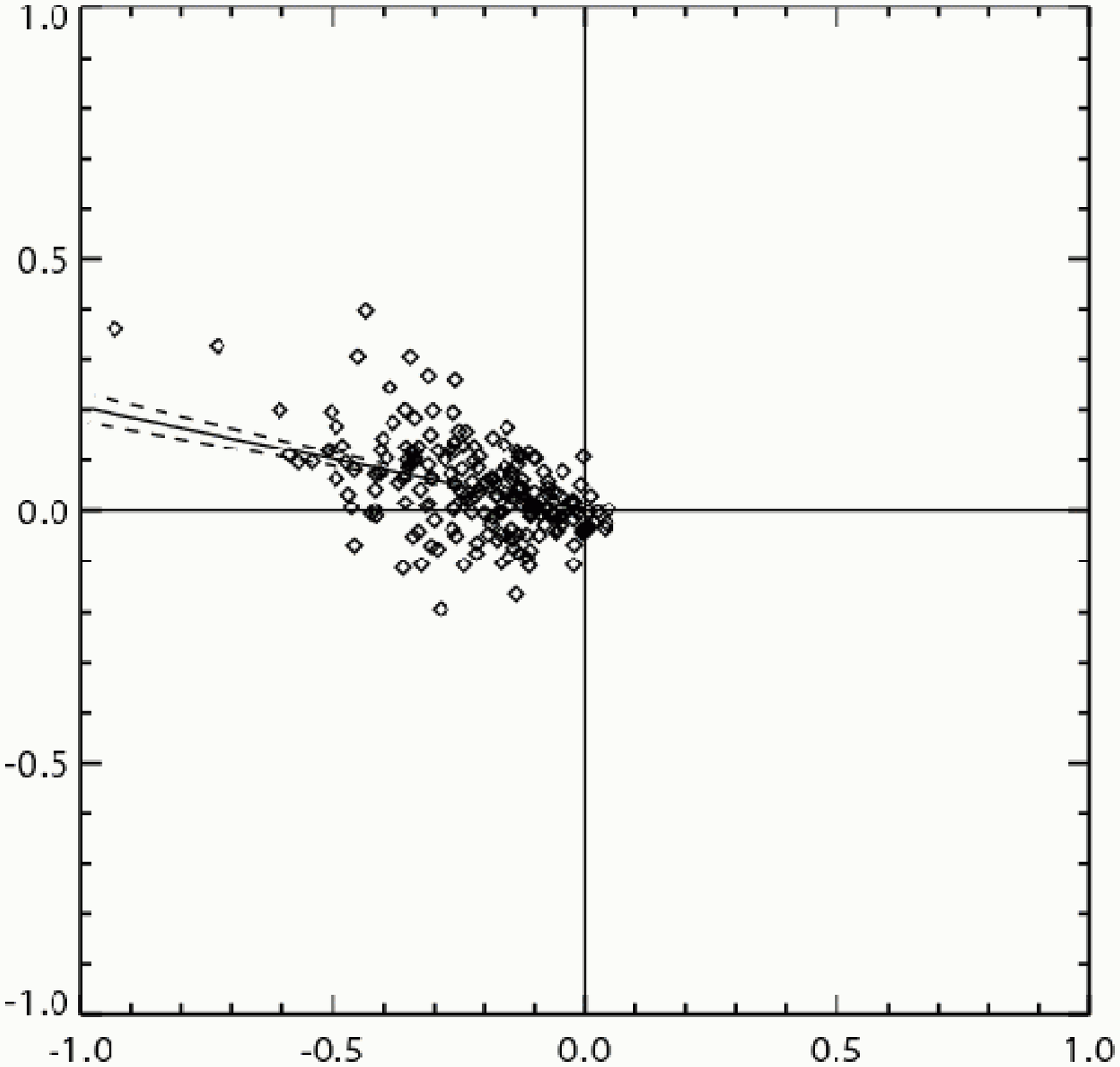}
\end{minipage}
\end{tabular}
\caption{\label{clp}Closure phase measurement for the star WR140 ($m_{H}=5.3$). The points represented on the plot are complex vectors normalised by the signal--to--noise. Each point is calculated using a single fringe measurement for each baseline (notably the N--S, N--W and W--S baselines, where the telescopes where positioned at N=35m, S=15m and W=10m). There are 200 points in the diagram. The vector represented in solid line is the average closure phase of the previous 200 vectors and the dashed line represents the error on the closure phase.  Finally the left panel represents a closure phase measurement in open--loop mode (FPT not active, closure~phase~=~$169.0 \pm 4.3$ deg) while the right panel is the closed--loop case (FPT active, closure~phase~=~$168.4 \pm 1.5$ deg).}
\end{figure}

\section{Results}

The FPT has been routinely in use at the IOTA interferometer since early 2002.
Fig.~\ref{fringetrack} shows the reduction of tracking residuals
operated by the algorithm. A reduction of about a factor 2 in the RMS change of the optical path, between data acquired with tracking switched off compared to data  with tracking on, can be
shown from the recorded data. This lies mostly in the low frequency
zone because of the limited bandwidth of the combined fringe-sensor,
fringe-actuator. For the IOTA interferometer this is a significant
improvement, since the high frequency phase noise is dealt within  post--processing
using the closure-phase information. 

\begin{figure}[!h]
\includegraphics[ scale=.6, angle=90]{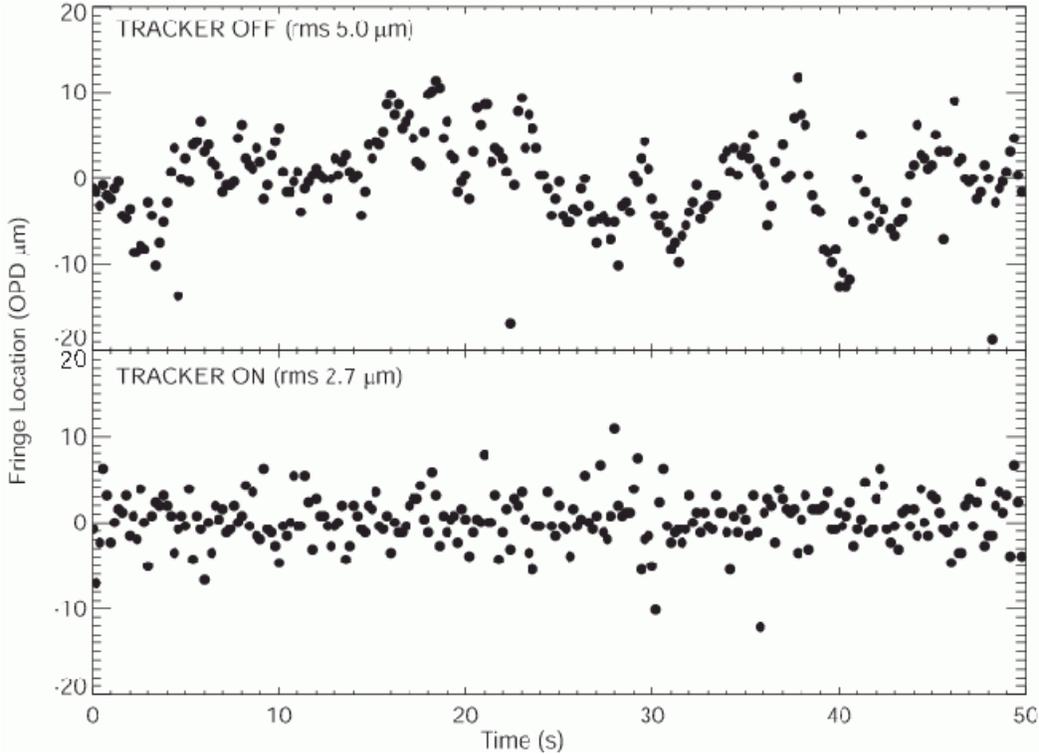}
\caption[The change of the OPD with
the FPT turned off.]{\label{fringetrack}The change of the OPD with
the FPT turned off (top) and on (bottom). This experiment was performed using a single, 21~m long baseline at 1.65~$\mu$m, during average seeing and no  baseline bootstrapping. The data was recorded the 6th of March 2002. We measure a factor 2 reduction in the RMS change in optical path.}
\end{figure}

It is in fact more important to
be able to maintain the fringe packet superposed and being able to
do so for faint sources rather than reduce the residuals of the OPD
to a smaller value. Fig.~\ref{clp} shows two measurements of closure phase. When the FPT is switched off the closure phase is $169.0 \pm 4.3 \deg$ but it is $168.4 \pm 1.5 \deg$ when the FPT is operating (a factor 3 error reduction for the closed--loop case).

We also show that the FPT is capable to track fringes on a 7.0~$m_H$ star with a SNR as low as 1.8 as shown in Fig.~\ref{snr}.

\begin{figure}[!h]
\includegraphics[  scale=.65,angle=90,origin=lB]{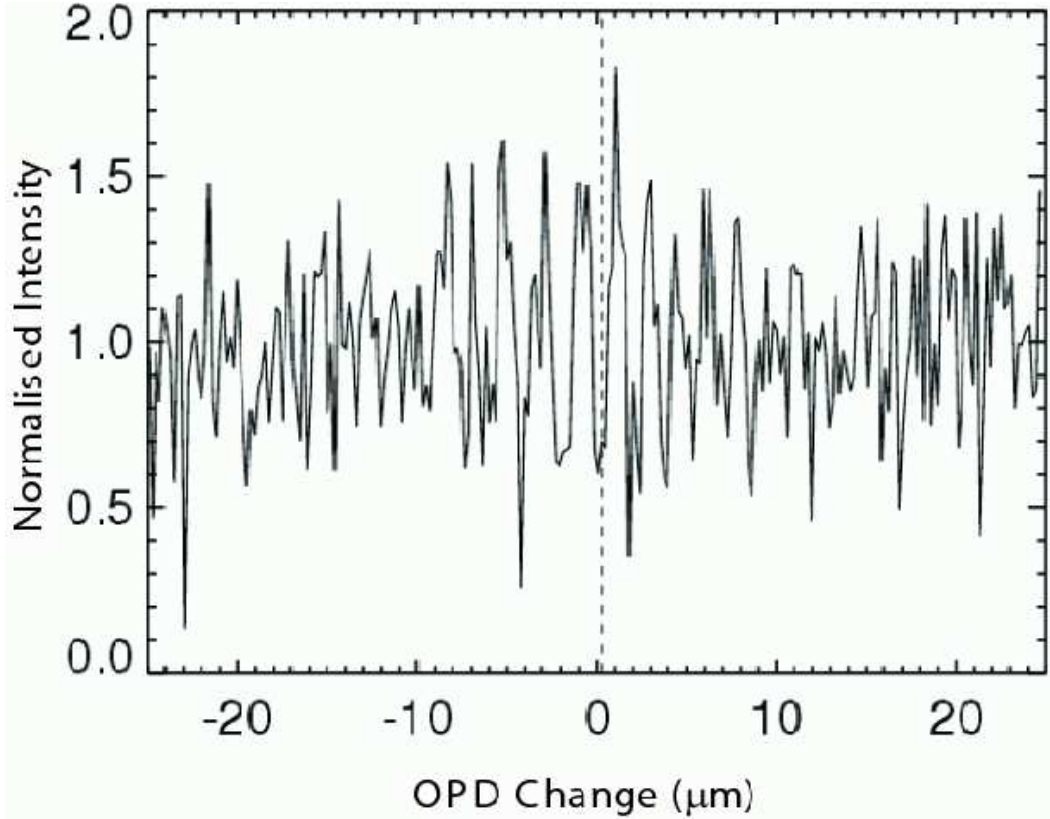}
\caption{\label{snr}Faint fringe observed on a 7.0~$m_H$ star, yielding a SNR of 1.8. The dashed line across the fringe  shows the expected position calculated by  the FTP.}
\end{figure}

\section{\label{sec:CONCLUSIONS}Conclusions}

We describe the fringe packet tracking system now routinely
in use at the IOTA interferometer. The packet tracker uses existing
hardware to perform its functions, notably the infrared camera used for data acquisition, as a
fringe sensor and the fringe scanning platforms combined with the
delay lines, for correcting the optical path. It is based on an
algorithm which exploits the wavelength-dependent information contained
in the fringe packet, which is extracted using double Fourier interferometry.
The phase of the fringes obtained for different frequencies is used
for calculating the group delay of the fringe packet. Since we use
baseline bootstrapping we are capable of blind-tracking fringes on
a baseline when their signal-to-noise is very low, provided that the
fringes have a good signal-to-noise on the other two baselines of
the interferometer. 

We use numerical simulation to model the case of
a fringe packet outside the scanning range, operated by the algorithm,
when only the side lobes of the fringe packet are visible.
We also show that the fringe packet tracker
algorithm effectively reduces the slowly varying atmospheric and instrumental-induced
additional path, leaving the fast phase noise to be dealt with post--processing. Moreover the FPT delivers about a factor 3 reduction  of the error on the closure--phase measurement and has been able so far to track fringes with SNR as low as 1.8.

\section{Acknowledgments}
This research was made possible thanks to a Smithsonian Predoctoral Fellowship  and a Michelson Postdoctoral Fellowship awarded to E. Pedretti.
The IONIC project is a collaboration among the Laboratoire d'Astrophysique
de Grenoble (LAOG), Laboratoire d'Electromagnetisme Microondes et
Optoelectronique (LEMO), and also CEA-LETI and IMEP, Grenoble, France.
The IONIC project is funded in France by the Centre National de Recherche
Scientifique and Centre National d'Etudes Spatiales. The work and
the fringe packet tracker were supported in part by grant NAG5-4900
from NASA, by grants AST-0138303 from the NSF, and by funds from the
Smithsonian Institution. This research has made use of NASA's Astrophysics
Data System Bibliographic Services.


\end{document}